\documentclass[prx,aps, amsmath,amssymb,floatfix,superscriptaddress,citeautoscript, twocolumn]{revtex4-1}
\usepackage{graphicx}
\usepackage{dcolumn}
\usepackage{bm}
\usepackage{amsmath}
\usepackage{amssymb}
\usepackage{amssymb, verbatim}
\usepackage[normal]{subfigure}
\usepackage{hyperref}
\hypersetup{colorlinks=true}

\begin{document}

\title{Quantum charge fluctuations of a proximitized nanowire}

\author{Roman M. Lutchyn}
\affiliation{Station Q, Microsoft Research, Santa Barbara, California 93106-6105, USA}
\author{Karsten Flensberg}
\affiliation{Center for Quantum Devices and Station Q Copenhagen, Niels Bohr Institute, University of Copenhagen, DK-2100 Copenhagen, Denmark}
\author{Leonid I. Glazman}
\affiliation{Department of Physics, Yale University, New Haven, CT 06520, USA}
\date{\today}

\begin{abstract}
Motivated by recent experiment~\cite{albrecht2015}, we consider charging of a nanowire which is proximitized by a superconductor and connected to a normal-state lead by a single-channel junction. The charge $Q$ of the nanowire is controlled by gate voltage $e{\cal N}_g/C$. A finite conductance of the contact allows for quantum charge fluctuations, making the function $Q(\mathcal{N}_g)$ continuous. It depends on the relation between the superconducting gap $\Delta$ and the effective charging energy $E^*_C$. The latter is determined by the junction conductance, in addition to the geometrical capacitance of the proximitized nanowire. We investigate $Q(\mathcal{N}_g)$  at zero magnetic field $B$, and at fields exceeding the critical value $B_c$ corresponding to the topological phase transition~\cite{Lutchyn10, Oreg10}. Unlike the case of $\Delta = 0$, the function $Q(\mathcal{N}_g)$ is analytic even in the limit of negligible level spacing in the nanowire. At $B=0$ and $\Delta>E^*_C$, the maxima of $dQ/d\mathcal{N}_g$ are smeared by $2e$-fluctuations described by a single-channel ``charge Kondo'' physics, while the $B=0$, $\Delta<E^*_C$ case is described by a crossover between the Kondo and mixed-valence regimes of the Anderson impurity model. In the topological phase, $Q(\mathcal{N}_g)$ is analytic function of the gate voltage with $e$-periodic steps. In the weak tunneling limit,  $dQ/d\mathcal{N}_g$ has peaks corresponding to Breit-Wigner resonances, whereas in the strong tunneling limit ({\it i.e.}, small reflection amplitude $r$) these resonances are broadened, and $dQ/d\mathcal{N}_g-e \propto r\cos(2\pi \mathcal{N}_g)$.
\end{abstract}
\maketitle

\section{Introduction}

The effect of BCS pairing on the energy spectrum of a system with fixed number of particles was first considered in the context of nuclear physics~\cite{Bohr1958} in an attempt to explain the oscillations in the excitation energy of a nucleus with the parity of nucleons number.  Oscillations of the nuclear mass captured by the phenomenological Weisz\"acker mass formula have the same origin~\cite{RevModPhys1958}.
A similar set of phenomena emerged in condensed matter physics in 90s, within the study of the Coulomb blockade in small ``islands'' of superconducting metals~\cite{Glazman'94, vonDelft'01}.  The charge $Q$ of a ``Coulomb island'' connected to a conducting lead by a low-conductance junction varies in steps upon variation of a gate voltage $V_g$ applied to the island via gate capacitor of capacitance $C_g$. In case of a normal-state island, the steps are of height $e$ and $e$-periodic as a function of the induced charge $e{\mathcal N}_g=C_g V_g$. Introduction of the $s$-wave superconductivity in the island changes the periodicity to $2e$. There is one step of height $2e$ within a period if the superconducting gap $\Delta$ exceeds the charging energy $E_C$. Upon reducing $\Delta$, each of the $2e$-steps split in two $e$-steps separated by a distance $[(E_C-\Delta)/E_C]e$; one returns to $e$-periodicity once gap is fully suppressed ($\Delta=0$)~\cite{Glazman'94}.

The majority of experiments addressing electron number parity phenomena in superconducting islands was performed on devices made of a conventional superconductor (Al) coupled to electrodes by junctions containing oxide barriers~\cite{Lafarge1993, Lafarge'93, Bouchiat1998, Lehnert'03}. Such barriers are not tunable, and typically have low transmission coefficients and large area (in units set by the Fermi wavelength of a typical superconductor). Most of experiments were fit well with theories\cite{Hekking'93, Houzet2005} considering the limit of small normal-state conductance of a junction, $G\ll G_q\equiv e^2/h$, in which case quantum fluctuations of charge are negligible.

\begin{figure}
\includegraphics[width=3.5in]{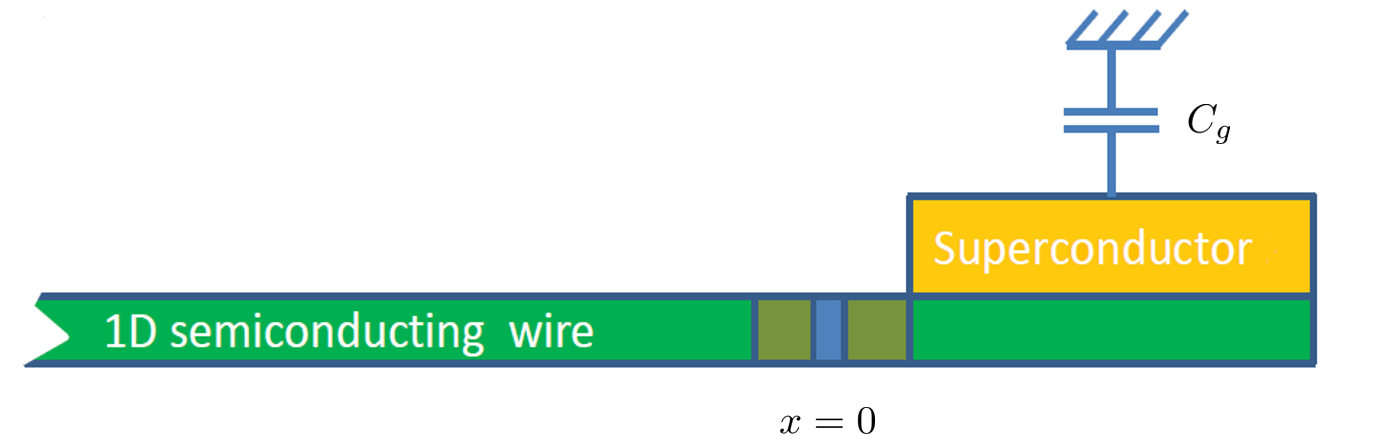}
\caption{Schematic plot of the device. A one-dimensional semiconductor nanowire is proximitized in the region $x >0 $. The barrier at $x=0$ can be electrostatically controlled. The charge of the proximitized nanowire can be controlled by changing the gate voltage $V_g$ applied through the capacitor $C_g$.}
\label{fig:device}
\end{figure}

Experiments with proximitized nanowires~\cite{Mourik2012,Das2012, Deng2012, Fink2012, Churchill2013, Deng_arxiv2014, Krogstrup2015, Higgi2015, Sherman'16} have opened a new aspect of the electron number parity phenomena. In the new devices~\cite{albrecht2015}, the fairly large induced gap values (comparable to those in superconducting aluminum) coexist with a large Fermi wavelength, so a junction may carry a single electronic mode (i.e. transverse area of the junction is of the order of Fermi wavelength), see Fig. \ref{fig:device}. Moreover, the junctions are gate-controlled, and their conductance can be tuned continuously, so that the dimensionless conductance $g$ may change from $g\ll 1$ to values approaching the unitary limit ($g$ is conductance per spin per channel measured in units of $e^2/h$). That leads to a substantial role of the quantum charge fluctuations. The quasiparticle gap $\Delta(B)$ can be suppressed by an applied magnetic field, and vanishes at the critical value $B=B_c$, signaling the emergence of a topologically-nontrivial state, carrying Majorana zero-energy modes~\cite{Sau10, Alicea10, Lutchyn10, Oreg10}. These zero-energy states may admit one electron. It makes the charge staircase to consist of equally-spaced $e$-steps, similar to the normal-state Coulomb blockade, despite the presence of a finite superconducting gap.

The goal of this work is two-fold. First, we elucidate the manifestation of the Majorana states in the charging effect by establishing the difference between the behaviors of a ``Majorana Coulomb island'' and the normal-state one. Second, we extend the theory of Coulomb blockade in superconducting islands to the case of single-channel, high-transmission junctions. Here we address both cases of $s$-wave and $p$-wave superconductivity in the island.

The paper is organized as follows. We begin with the spinful model and consider in Sec.~\ref{sec:spinful} Coulomb blockade of a proximitized nanowire in the weak and strong tunneling limits. In Sec.~\ref{sec:spinless}, we introduce a spinless model for a proximitized nanowire (i.e. Majorana nanowire) and discuss charging effect as a function of the junction conductance $g$ and the quasiparticle gap $\Delta_P$. Finally, we conclude with the qualitative summary of our results in Sec. \ref{sec:conclusions}.

\section{Charge discreteness effect in the topologically trivial state (zero magnetic field)}\label{sec:spinful}

\subsection{Coulomb blockade in the weak-tunneling limit}\label{sec:weak}

In the weak tunneling limit $g\ll 1$, it is convenient to use the formalism of tunneling Hamiltonian (hereafter $\hbar=1$):
\begin{align}
H&=\sum_{k,\sigma}\xi_{k}a_{k\sigma}^\dagger a_{k\sigma}
+t\sum_{k,p,\sigma}\left( a_{k\sigma}^\dagger a_{p\sigma}+h.c.\right)+H_{\rm BCS}\nonumber\\
&+E_C(Q/e-{\mathcal N}_g)^2\,.
\label{tunnH}
\end{align}
Here the first term describes the normal-state conductor, the second term is the Hamiltonian of the tunnel junction. The tunneling constant can be related to the dimensionless conductance $g={4\pi^2\nu|t|^2}/\delta$,
where $1/\delta$ and $\nu$ are the average density of levels in the proximitized nanowire and the density of states in the normal lead per spin, respectively. The third term is the particle-conserving form of the BCS Hamiltonian~\cite{tinkham1}, and the last term describes the charging effect, $E_C=e^2/2C_\Sigma$, where $C_\Sigma$ is the total capacitance of the proximitized wire to all other electrodes including gate). The labels $k$ and $p$ enumerate orbital states in the normal lead and in the proximitized wire, respectively, and tunneling through a potential barrier preserves the Kramers pair states labeled by $\sigma$. The charge $Q(\mathcal{N}_g)$ can be expressed in terms of the ground-state average
\begin{align}\label{eq:charge}
 Q(\mathcal{N}_g)=e \mathcal{N}_g+q(\mathcal{N}_g)=e \mathcal{N}_g-\frac{e}{2E_C}\left \langle\frac{\partial H}{\partial{\cal N}_g} \right \rangle,
\end{align}
where $q(\mathcal{N}_g)$ is the so-called reduced charge.

In the case of a large BCS gap, $\Delta>E_C$, the low-energy subspace includes the particle-hole excitations of the normal-state conductor and the two possibly degenerate charge states of the wire. The degeneracy occurs at odd-integer values of ${\cal N}_g=2n+1$, and the involved charge states of the proximitized wire are $|2n\rangle$ and $|2(n+1)\rangle$. Tunneling removes the degeneracy. To explore the resulting state, it is convenient to reduce the problem to the low-energy subspace, $E\lesssim E_C\ll\Delta$. By performing a Schrieffer-Wolff transformation on the Hamiltonian~(\ref{tunnH}) at ${\cal N}_g=2n+1$, one finds
\begin{align}
H\!&=\!\sum_{k,\sigma}\xi_{k}a_{k\sigma}^\dagger a_{k\sigma}
\!+\!\frac{t_A}{\Omega}\!\sum_{k_1,k_2,\sigma}\!\!\!\sigma a_{k_1\sigma}^\dagger a_{k_2-\sigma}^\dagger
|2n\rangle\langle 2n\!+\!2|+h.c.\nonumber\\
&\!+\!\frac{r}{2\Omega}\sum_{k_1,k_2,\sigma}\!\!\! a_{k_1\sigma}^\dagger a_{k_2\sigma}
\left(|2n\rangle\langle 2n|-|2n\!+\!2\rangle\langle 2n\!+\!2|\right)\,
\label{eq:AH}
\end{align}
with $\Omega$ being the volume of the lead. The coefficients $t_A$ and $r$ correspond to Andreev and normal scattering, respectively. The amplitude for these processes were evaluated in Refs. \cite{Hekking'93, Garate'11}
\begin{align}
|\nu t_A|=&\frac{2\nu|t|^2}{\delta} \frac{1}{\sqrt{1-x^2}}\arctan\sqrt{\frac{1+x}{1-x}},\\
|\nu r|=& \frac{2\nu|t|^2}{\delta} \left(\frac{x}{1-x^2}\arctan \sqrt{\frac{1+x}{1-x}}\right.\nonumber\\
&\left.+\frac{3 x}{1-9x^2}\arctan \sqrt{\frac{1-3x}{1+3x}}\right),
\label{tA}
\end{align}
where $x=E_C/\Delta$. In Eq.\eqref{eq:AH} we omitted insignificant, non-singular correction $\propto t^2$ to the charging energy $E_C$ and a non-singular potential scattering term $k_1\sigma\to k_2\sigma$ of the same order proportional to $|2n\rangle\langle 2n|+|2n+2\rangle\langle 2n+2|$.

The Andreev reflection term proportional to $t_A$ in Eq.~\eqref{eq:AH} is similar to the spin-flip term in the anisotropic Kondo model, except that the role of the local spin is played by the charge of the proximitized wire, projected onto the subspace of states $|2n\rangle$ and $|2n+2\rangle$. The ``Ising component'' in Eq.~(\ref{eq:AH}), proportional to $r$, corresponds to normal scattering processes. By performing a particle-hole transformation for one of the spin components (i.e. $\psi_{\downarrow}(x)\rightarrow \psi^{\dag}_{\downarrow}(x)$), one can map the Hamiltonian~\eqref{eq:AH} to the conventional (spin) Kondo Hamiltonian. Therefore, we will refer to the quantum mechanics problem defined by Eq.~(\ref{eq:AH}) as the charge Kondo problem~\cite{Glazman'90, Matveev'91, Matveev'95}. The corresponding Renormalization Group (RG) equations for the coupling constants $t_A$ and $r$ are given by
\begin{align}
\frac{d \tilde{t}_A}{dl}&=2 \tilde{t}_A \tilde{r},\\
\frac{d \tilde{r}}{dl}&=2 \tilde{t}_A^2,
\end{align}
where $l=\ln(E_C/D)$ with $D$ being the running cut-off. Here tilde denotes rescaled constants, i.e. $\tilde{r}\equiv \nu r$.
In the limit $E_C \ll \Delta$, the initial values for $\tilde{t}_A(0)\equiv \nu t_A$ and $\tilde{r}(0)\equiv \nu r$ correspond to $\tilde{t}_A(0) \gg \tilde{r}(0)$, and the solution of the RG equations reads
\begin{align}
\tilde{r}(l)=-\kappa\cot\left(2\kappa l -\alpha\right),\\
\tilde{t}_A(l)=-\kappa\csc\left(2\kappa l -\alpha\right),
\end{align}
where $\kappa=\sqrt{\tilde{t}_A(0)^2-\tilde{r}(0)^2}$ and $\alpha=\arctan\left(\frac{\kappa}{\tilde{r}(0)}\right)$. The strong coupling fixed point is reached when ${\rm max} \{\tilde{r}(l), \tilde{t}_A(l)\}\sim 1$ defining the charge Kondo energy scale
\begin{equation}
T_K\sim E_C\exp\left(-\frac{\pi}{4|\nu t_A|}\right)\equiv E_C \exp\left(-\frac{\pi^2}{g}\right), \,
\label{TK}
\end{equation}
where $g \ll 1$ is the dimensionless conductance in units of $e^2/h$.

\begin{figure}
\includegraphics[height=2.1in]{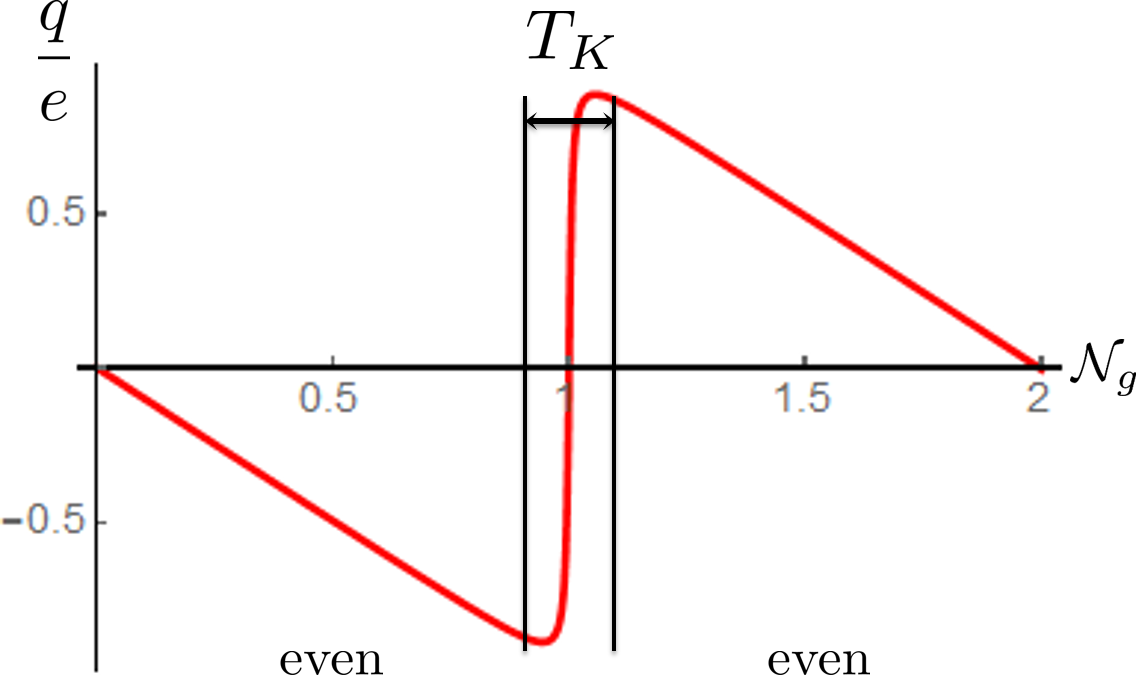}
\caption{Reduced charge of a superconducting nanowire as a function of the dimensionless gate voltage $\mathcal{N}_g$ for $\Delta > E_C$. Smearing of the charge steps is controlled by the Kondo temperature scale $T_K$~\eqref{TK}.}
\label{fig:chargeKondo}
\end{figure}

The scale $T_K$ defines the smearing of the charge steps, see Fig.~\ref{fig:chargeKondo}. Since the low-energy behavior of the ``conventional'' single-channel Kondo model is described by the Fermi liquid, it yields an analytic dependence of magnetization on the magnetic field~\cite{Yosida}. Similarly, in the charge Kondo problem the dependence of $Q$ on ${\cal N}_g$ is analytic; the maximal differential capacitance (corresponding to odd-integer values of ${\cal N}_g$) is $dQ/d{\cal N}_g\sim e \exp(\pi^2/g)$.

Next we consider a larger charging energy, $E_C>\Delta$.  At $g\to 0$, the transition between ``even'' and ``odd'' charge plateaus occurs at gate voltages
\begin{equation}\label{eq:shift}
{\cal N}_g^*=2n+\frac{E_C+\Delta}{2E_C}\,
\end{equation}
with $n \in \mathbf{Z}$. At these points, the finite-order perturbation theory in $t$ for $Q({\cal N}_g)$ is divergent. Summation of the most divergent terms can be performed~\cite{Houzet2005} in the limit of a large number of channels in the junction $N_{\rm ch} \gg 1$. In the leading order in $1/N_{\rm ch}$, one finds~\cite{Houzet2005}
\begin{align}
Q=(2n+1)e-e\theta({\cal N}_g^*-{\cal N}_g)f\left(\frac{{\cal N}_g^*-{\cal N}_g}{\delta{\cal N}_g}\right) ,
\label{qpt}
\end{align}
where $f(x)=1-\frac{1}{\sqrt{1+x}}$ and $\delta{\cal N}_g=\left(\frac{g}{8\pi}\right)^2\frac{\Delta}{E_C}$. The function $Q({\mathcal N}_g)$ defined by Eq.~(\ref{qpt}) is continuous but not analytic: in the limit $N_{\rm ch}\to\infty$ a quantum phase transition occurs at ${\mathcal N}_g={\mathcal N}_g^*$. On the odd side of the transition, ${\mathcal N}_g>{\mathcal N}_g^*$, a bound state for a single electron near the junction is formed in the proximitized wire; this state is doubly-degenerate in the electron spin. The spin degeneracy is in fact a consequence of the unphysical limit $N_{\rm ch}\to\infty$. At a finite number of channels, the residual tunneling between the localized state and the continuum in the normal lead leads to an exchange interaction between the bound electron and the Fermi sea of the normal lead (this is similar to the effect of hybridization in the Anderson impurity model). The exchange removes the spin degeneracy, as in the conventional Kondo effect. As a result, the system exhibits a crossover rather than a transition at ${\mathcal N}_g={\mathcal N}_g^*$, similar to the crossover between the Kondo and mixed-valence regimes of the Anderson impurity model~\cite{Haldane'78}. We will demonstrate the absence of the phase transition in the next section using the bosonization scheme, valid for a single-channel case.

\begin{figure}
\includegraphics[height=2.1in]{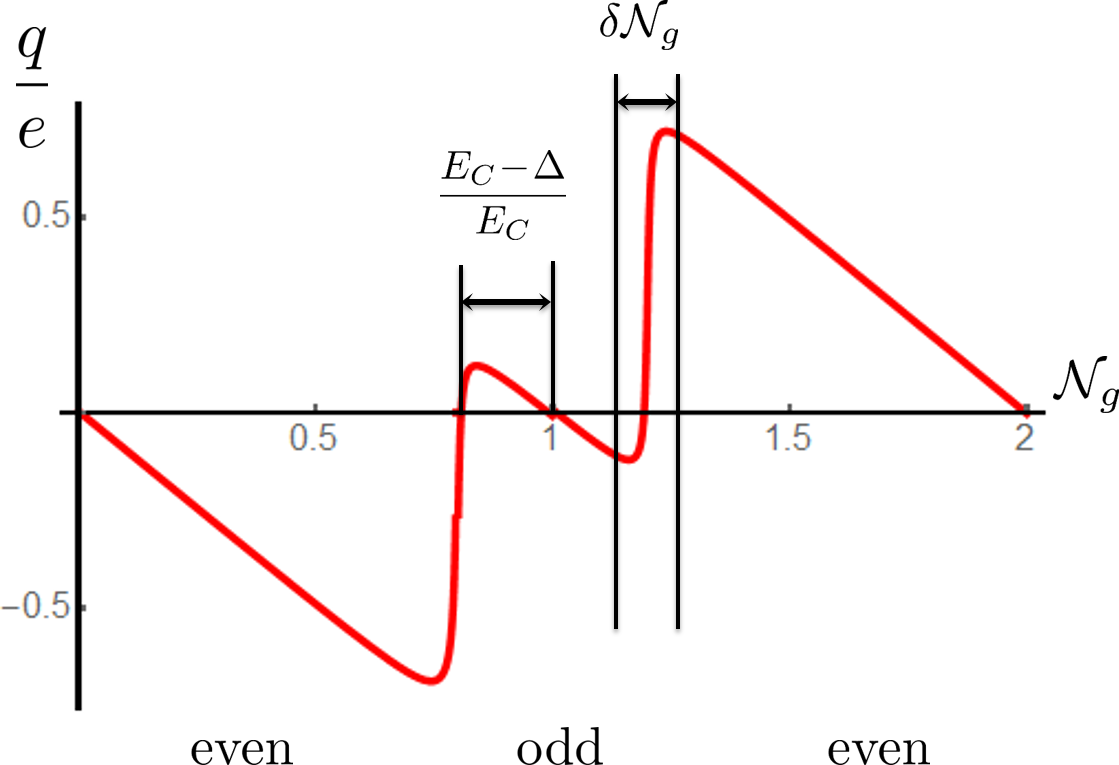}
\caption{Reduced charge of a superconducting nanowire as a function of the dimensionless gate voltage $\mathcal{N}_g$ for $E_C>\Delta$. The left transition point occurs at $\mathcal{N}_g=\mathcal{N}_g^*$, see Eq.~\eqref{eq:shift}. The width of the steps is $\delta{\cal N}_g=\left(\frac{g}{8\pi}\right)^2\frac{\Delta}{E_C}$, see Eq.~\eqref{qpt}.}
\label{fig:chargehouzet}
\end{figure}


\subsection{Coulomb blockade in the strong-tunneling limit}\label{sec:strong}

In the strong tunneling limit corresponding to a small reflection amplitude $r$ at the junction (i.e. $r \ll 1$), it is more convenient~\cite{Flensberg'93,Matveev'95} to analyze the charging effect in terms of bosonic variables~\cite{giamarchi},
\begin{align}
\psi_{R/L,\sigma}=\frac{1}{\sqrt{2\pi a}}\exp\left(-\frac{i}{\sqrt{2}}\left[\pm \phi_{\rho}-\theta_{\rho}+\sigma (\pm \phi_{\sigma}-\theta_{\sigma})\right]\right)
\end{align}
with $a$ being the ultraviolet cutoff length scale. The low-energy excitations in Coulomb island ({\it i.e.}, a quantum dot with a dense single-particle spectrum) connected to a bulk conductor by a single-mode junction are described by an effective one-dimensional model
\begin{align}\label{eq:H_normal}
H&=H_W+H_P+H_C+H_B\,,\\
H_W&\!=\!\sum_{\lambda=\rho, \sigma} \frac{v}{2\pi}\int_{-\infty}^{L}\!\!dx \left (\partial_x \phi_\lambda)^2\!+\! (\partial_x \theta_\lambda)^2\right]\,,\\
H_C&=  E_{C}\left(\frac{\sqrt{2}}{\pi}{\phi}_{\rho}(0)-\mathcal {N}_g\right)^2\,,\\
H_B&= - D r  \cos \sqrt{2} {\phi}_\rho(0) \cos \sqrt{2} {\phi}_\sigma(0)\,.
\end{align}
Here $E_C$ is the bare charging energy defined by the electrostatic environment of the proximitized wire; $H_B$ represents normal backscattering at the junction. To describe the proximity-induced gap in the spectrum, we add the following term to Eq.~\eqref{eq:H_normal}:
\begin{equation}
H_P=-\frac{\Delta}{2\pi a} \int_{0}^{L} dx \cos \sqrt 2 {\theta}_\rho \cos \sqrt 2{\phi}_\sigma.
\label{eq:H_Delta}
\end{equation}
We implicitly assume that the quasiparticle gap in the source of the superconducting proximity effect ({\it e.g.}, aluminium) $\Delta_{\rm Al}$ is large, $\Delta_{\rm Al} \gg \Delta, E_C$, so that the corresponding excitations are absent at energies below $\Delta_{\rm Al}$. Therefore, the bandwidth $D\sim v/a$ in the effective Hamiltonian~\eqref{eq:H_normal},\eqref{eq:H_Delta} should satisfy the condition $D\ll \Delta_{\rm Al}$. The Hamiltonian $H_W+H_P$ is a bosonized version of a BCS model in one dimension; the values of velocity $v$ and gap $\Delta$ should be properly tuned to reproduce the quasiparticle spectrum~\cite{Stanescu'11} of a proximitized wire. Hereafter we will also assume that the proximitized nanowire is long such that the corresponding level spacing in that spectrum is negligible ($v/L \rightarrow 0$). In the bosonic representation, a quasiparticle can be viewed as a kink in the spin field ${\phi}_\sigma$ which costs energy $\Delta$.

We note from the outset that charge discreteness effects vanish at $r=0$, regardless of the presence of the pairing term~\eqref{eq:H_Delta}. Indeed, at $r=0$ we may exclude the dependence of the Hamiltonian \eqref{eq:H_normal} -\eqref{eq:H_Delta} on $\mathcal {N}_g$ by performing a shift transformation $\phi_{\rho}(0) \rightarrow \phi_{\rho}(0)+\pi \mathcal {N}_g/\sqrt{2}$, since the corresponding shift operator commutes with $H_P$.

When $r\neq 0$, the aforementioned transformation moves the gate-voltage dependence from the charging energy $H_C$ to the backscattering term $H_B$ of the Hamiltonian \eqref{eq:H_normal}, without affecting other terms:
\begin{align}\label{eq:H_normal1}
&H=H_0+H_B,\,\, H_0=\!\sum_{\lambda=\rho, \sigma} \frac{v}{2\pi}\int_{-\infty}^{L}\!\!dx \left (\partial_x \phi_\lambda)^2\!+\! (\partial_x \theta_\lambda)^2\right]
\nonumber\\
&-\frac{\Delta}{2\pi a} \int_{0}^{L} dx \cos \sqrt 2 {\theta}_\rho \cos \sqrt 2{\phi}_\sigma+E_{C}\left(\frac{\sqrt{2}}{\pi}{\phi}_{\rho}(0)\right)^2\,,
\nonumber\\
&H_B=\!- D r  \cos\left(\sqrt{2} {\phi}_\rho(0)-\pi\mathcal {N}_g\right) \cos \sqrt{2} {\phi}_\sigma(0)\,.
\end{align}
At small $|r|$, we may investigate the gate-voltage-dependent part of the ground-state energy, $\delta E_{\rm GS}(\mathcal{N}_g)$ by developing a perturbation theory in $r$. Expanding up to the second order in $r$, one finds that $\delta E_{\rm GS}(\mathcal{N}_g)=\delta E_{\rm GS}^{(1)}+\delta E_{\rm GS}^{(2)}$ with
\begin{align}
\delta E_{\rm GS}^{(1)}&=\langle H_B\rangle_{H_0}
\label{eq:firstorder}\\
&=-Dr\left\langle\cos\sqrt{2} {\phi}_\rho(0,\tau) \cos \sqrt{2} {\phi}_\sigma(0,\tau)\right\rangle\cos\left(\pi\mathcal {N}_g\right)\,,
\nonumber\\
\delta E_{\rm GS}^{(2)}&=\int_{0}^{\beta} d\tau \langle H_B(\tau)H_B(0)\rangle_{H_0}
\label{eq:secondorder}\\
&\!\!=\!-D^2r^2\! \cos^2\left(\pi\mathcal {N}_g\right)\!\! \int_{0}^{\beta}\! d \tau
\left \langle e^{i\sqrt{2}[{\phi}_\rho(0,\tau) + {\phi}_\rho(0,0)] } \right \rangle
\nonumber\\
&\left \langle \cos \sqrt{2} {\phi}_\sigma(0,\tau)\cos \sqrt{2} {\phi}_\sigma(0,0)
\right\rangle\,
\nonumber
\end{align}
with $\tau$ and $\beta$ being the imaginary time and inverse temperature. Henceforth we consider the zero temperature limit $\beta \rightarrow \infty$. The main difference of Eqs.~(\ref{eq:firstorder}), (\ref{eq:secondorder}) with respect to the normal Coulomb island case~\cite{Matveev'95} is a finite value of $\delta E_{\rm GS}^{(1)}(\mathcal {N}_g)$ which, in fact, ensures the $2e$-periodicity of the observable quantities. It emerges due to the presence of the superconducting gap which suppresses the spin-mode fluctuations.

To estimate the ground-state average $\langle\dots\rangle$ entering $\delta E_{\rm GS}^{(1)}(\mathcal {N}_g)$, we divide the quantum fluctuations of the fields ${\phi}_\rho$ and ${\phi}_\sigma$ into three regions corresponding to energy intervals (a) $\varepsilon>\max(\Delta,E_C)$, (b) $\max(\Delta,E_C)>\varepsilon>\min(\Delta,E_C)$, and (c) $\min(\Delta,E_C)>\varepsilon$. In each of the regions we simplify the Hamiltonian $H_0$ to a quadratic form, neglecting $H_C$ and $H_P$ in the interval (a), and keeping their quadratic expansions at energies below $E_C$ and $\Delta$, respectively. This approximation allows us to factorize the averages of $\cos\sqrt{2} {\phi}_\rho$ and $\cos \sqrt{2} {\phi}_\sigma$ and estimate $\langle\cos \sqrt{2} {\phi}_\sigma\rangle\sim\sqrt{\Delta/D}$, irrespective to the relation between $\Delta$ and $E_C$. The estimate of $\langle\cos \sqrt{2} {\phi}_\rho\rangle$, however, depends on which of the two scales is the largest one.

We consider first the case $\Delta\gg E_C$. The presence of the spectral gap resulting in pinning of the field ${\theta}_\rho(x)$ at $x>0$ by the pairing energy Eq.~(\ref{eq:H_Delta}) allow us to consider fluctuations of the field ${\phi}_\rho(x)$ only at $x<0$. Using the continuity of the fields and the free-field equation of motion for $x<0$, one finds the following boundary condition $\partial_x{\phi}_\rho(0^-)=0$. The singled out three energy intervals yield three factors in the average over the charge density mode, $\langle\cos \sqrt{2} {\phi}_\rho\rangle\sim\sqrt{\Delta/D}\cdot(E_C/\Delta)\cdot 1$. Collecting all the factors, we arrive at
\begin{equation}\label{eq:charging_spinful1}
\delta E_{\rm GS}(\mathcal{N}_g)\sim - E_C r \cos (\pi \mathcal{N}_g)\,,
\end{equation}
where we neglected a smaller, $\sim E_Cr^2$ second-order in $r$ contribution~\eqref{eq:secondorder}. We may associate the energy scale in Eq.\eqref{eq:charging_spinful1} with the effective charging energy $E_C^*=E_C r$.

In the case of a smaller gap, $E_C\gg\Delta$, the charge density fluctuations at $x=0$ are pinned by the Coulomb energy $H_C$ at energies $\varepsilon\sim E_C$, which are much higher than $\Delta$. The pinning by $H_C$ affects the three factors entering the average of the charge density fluctuations: $\langle\cos \sqrt{2}{\phi}_\rho\rangle\sim\sqrt{E_C/D}\cdot 1\cdot 1$, resulting in
\begin{equation}\label{eq:charging_spinful2}
\delta E_{\rm GS}^{(1)}(\mathcal{N}_g)\sim - \sqrt{E_C\Delta} r \cos (\pi \mathcal{N}_g).
\end{equation}
In evaluation of the second-order in $r$ term, we may follow Ref.~\cite{Matveev'95} in noticing that the long-time asymptote of the integrand of~\eqref{eq:secondorder} yields a logarithmically-large contribution. In our case, the corresponding imaginary time interval runs from $1/E_C$ to $1/\Delta$. Thus, the logarithmic divergence is cutoff by the superconducting gap $\Delta$. As a result, we find
\begin{equation}\label{eq:charging_spinful3}
\delta E_{\rm GS}^{(2)}(\mathcal{N}_g)\sim - E_C r^2 \ln\frac{E_C}{\Delta} \cos^2 (\pi \mathcal{N}_g).
\end{equation}
Thus, the charge of the nanowire $Q(\mathcal{N}_g)$, calculated using $\delta E_{\rm GS}(\mathcal{N}_g)=\delta E_{\rm GS}^{(1)}+\delta E_{\rm GS}^{(2)}$, is a continuous and analytic function of $\mathcal{N}_g$. As long as $\Delta\gg E_Cr^2$, the second-order contribution is relatively small for the entire range of gate voltages.

Perturbation theory in $r$ breaks down at $\Delta\sim E_Cr^2$. Furthermore, in the limit $\Delta \ll E_Cr^2$ one expects a transition between even and odd-charge sectors, similar to the weak tunneling case as depicted in Fig.~\ref{fig:chargehouzet}. As explained in Sec.~\ref{sec:weak}, the even-odd transition for $N_{\rm ch} \rightarrow \infty$ is accompanied by the non-analytic behaviour of the ground-state energy, see Eq.~\eqref{qpt}. We now demonstrate, using a single-channel model, that quantum fluctuations smear out these non-analyticities and ultimately destroy the quantum phase transition. As a result, the function $Q(\mathcal {N}_g)$ remains analytic.

We now concentrate on the limit $\Delta \ll E_Cr^2$. The boundary term $H_B$  in the Hamiltonian~\eqref{eq:H_normal} tends to pin the field ${\phi}_\sigma(0)$ at a value which depends on the gate voltage $\mathcal{N}_g$. Indeed, due to the presence of the charging energy we may replace $H_B$ in~\eqref{eq:H_normal} by its average with respect to $\varphi_{\rho}(0)$:
\begin{equation}
H_B\rightarrow
\langle H_B\rangle_{_{\varphi_\rho}}\!\!
\sim -r\sqrt{DE_C} \cos\left(\pi\mathcal {N}_g\right) \cos \sqrt{2} {\phi}_\sigma(0).
\label{eq:H_sigma}
\end{equation}
The value of ${\phi}_\sigma(0)$ which minimizes the energy~\eqref{eq:H_sigma} depends on the sign of $\cos (\pi \mathcal{N}_g)$:
\begin{align}
\sqrt{2} {\phi}_\sigma(0)&=2\pi k \mbox{   for   }  |\mathcal{N}_g-2n|<1/2 \label{eq:phisigmabc1}, \\
\sqrt{2} {\phi}_\sigma(0)&=\pi (2k+1) \mbox{ for }  |\mathcal{N}_g-(2n+1)|<1/2.
\label{eq:phisigmabc2}
\end{align}
Here we implicitly assumed that $\sqrt{2}{\phi}_\sigma(L)$ is an even integer and $k, n \in \mathbf{Z}$. In the absence of superconducting pairing ($\Delta=0$), the two configurations of the spin mode that differ by the boundary condition, Eqs.~(\ref{eq:phisigmabc1}), (\ref{eq:phisigmabc2}), may have the same energy since the field ${\phi}_\sigma(x)$ is free to fluctuate in the region $x \in [0,L]$. However, with finite pairing $\Delta>0$, the field $\sqrt{2}{\phi}_\sigma (x)/\pi$ is pinned to an even integer in the bulk. This is compatible with the boundary condition in the even valley~\eqref{eq:phisigmabc1}, but not with the condition in the odd valley~\eqref{eq:phisigmabc2}. As a result, to the linear order in $\Delta$ the function $\delta E_{\rm GS}(\mathcal{N}_g)$ behaves differently in the even-charge and odd-charge domains of $\mathcal{N}_g$. In the former, we may use the $\Delta=0$ result derived in Ref.~\cite{Matveev'95}. However,  if the gate voltage belongs to an odd-charge domain $|\mathcal{N}_g-(2n+1)|<1/2$, there is a $\sqrt{2}\pi$-kink in the ground state of the system. Such a kink corresponds, as we already mentioned, to a quasiparticle in the proximitized  nanowire segment; it increases the ground state energy by $\Delta$. Thus, the ground state energy of the system in the corresponding limit reads
\begin{align}
&\delta E_{\rm GS}(\mathcal{N}_g)\sim
-r^2E_C\ln\left(\frac{1}{r^2\cos^2\pi\mathcal{N}_g}\right)
\cos^2\pi\mathcal{N}_g
\nonumber\\
&\mbox{   at   }  1/2-|\mathcal{N}_g-2n|\gg(\Delta/E_Cr^2)^{1/2}\,, \label{eq:eg_even} \\
&\delta E_{\rm GS}(\mathcal{N}_g)\sim
-r^2E_C\ln\left(\frac{1}{r^2\cos^2\pi\mathcal{N}_g}\right)
\cos^2\pi\mathcal{N}_g+\Delta
\nonumber\\
&\mbox{ at }  1/2-|\mathcal{N}_g-(2n+1)|\gg (\Delta/E_Cr^2)^{1/2}\,.
\label{eq:eg_odd}
\end{align}
In the regions of $\mathcal{N}_g$ excluded from Eqs.~(\ref{eq:eg_even}) and (\ref{eq:eg_odd}), perturbation theory in $\Delta$ breaks down.  The dependence $\delta E_{\rm GS}(\mathcal{N}_g)$ is expected to be continuous, $2e$-periodic, with maxima shifted from half-integer points $\mathcal{N}_g=n+1/2$ into the odd-charge domains by an amount $\sim (\Delta/E_Cr^2)^{1/2}$. In the following, we are not interested in the detailed dependence of $\delta E_{\rm GS}(\mathcal{N}_g)$ within the regions $|\mathcal{N}_g-1/2-n|\lesssim (\Delta/E_Cr^2)^{1/2}$. Instead, we will concentrate on the analytic properties of that dependence and show that the function $\delta E_{\rm GS}(\mathcal{N}_g)$ becomes analytic at the even-odd charge transitions, in contrast to Eq.~\eqref{eq:eg_odd}.

To investigate the even-odd transition, we run the RG procedure until $\widetilde{D}\sim\Delta$ and formulate a low-energy problem in which the charge density degrees of freedom are already frozen, and only the spin excitations with energy $\varepsilon\lesssim\Delta$ are accounted for. During the RG flow the coefficient of the backscattering term changes in a way that depends on the initial values and the gate voltage, and we introduce a function $\Gamma(\mathcal{N}_g)$ to parameterize this dependence. The corresponding imaginary-time action then takes the form
\begin{align}\label{eq:stationary}
&S=S_W+S_P+S_B,\\
&S_W=\frac{1}{2\pi}\int d \tau\int_{-\infty}^{L}\frac{dx}{v} \left[(\partial_\tau {\phi}_{\sigma})^2+ v^2(\partial_x {\phi}_{\sigma})^2\right],\\
&S_P=-\Delta\frac{\widetilde{D}}{v} \int d \tau \int_{0}^{L} dx \cos \sqrt 2 {\phi}_\sigma,\\
&S_B=   -\Gamma(\mathcal{N}_g) \int d \tau   \cos(\sqrt{2} {\phi}_\sigma(0)).
\end{align}
To understand the putative transition, we expand  $\Gamma(\mathcal{N}_g)
\approx \Gamma(\mathcal{N}^*_g)+\widetilde{\Gamma} (\mathcal{N}_g-\mathcal{N}_g^*)$ near the charge degeneracy point $\mathcal{N}_g^*$. Based on Eq.~\eqref{eq:eg_odd}, we estimate that $\mathcal{N}_g^*-1/2\sim (\Delta/E_Cr^2)^{1/2}$ and $\widetilde{\Gamma}\sim\sqrt{E_C\Delta}r$. We take $\widetilde{D}\sim\Delta$ for the bandwidth of excitations of the field ${\phi}_{\sigma}(x)$.

\begin{figure}
\includegraphics[width=3.4in]{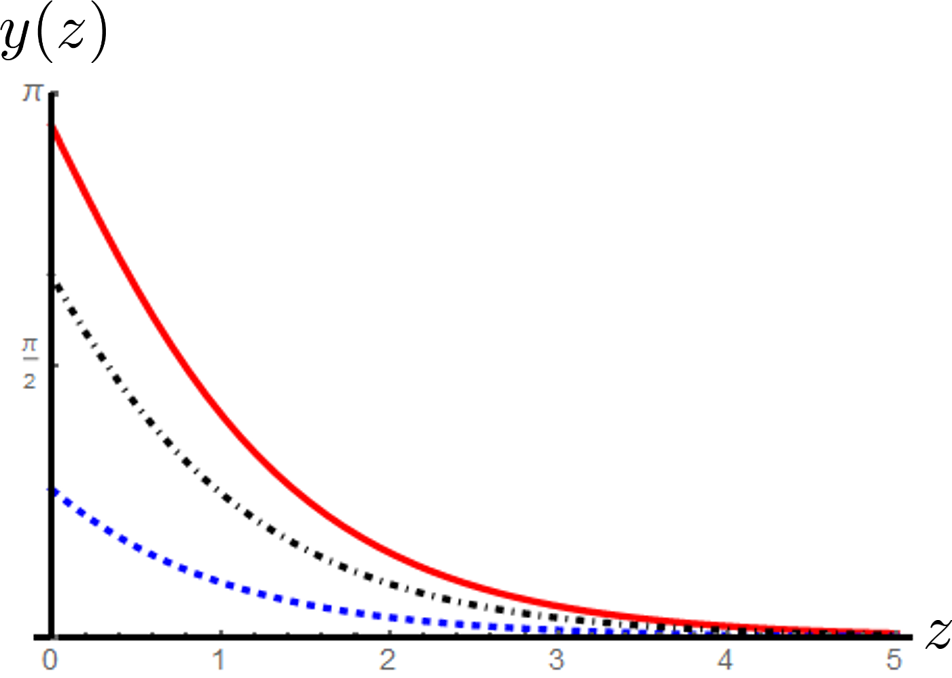}
\caption{Inhomogeneous classical solution $y(z)$ for an odd-charge configuration as a function of detuning from the putative degeneracy point $\eta$. Here solid (red), dot-dashed (black) and dashed (blue) lines correspond to $\eta=10, 1, 0.1$, respectively.}
\label{fig:csoliton}
\end{figure}

We now introduce dimensionless variables $x\rightarrow z \xi$, $\tau \rightarrow \tilde{\tau} \xi/v$ with $\xi\sim v/\Delta$ and rewrite the effective action \eqref{eq:stationary} as
\begin{widetext}
\begin{align}\label{eq:app:action}
S=\frac{1}{2\pi} \int d\tilde{\tau} \int_{-\infty}^{\tilde{L}}dz\left\{\frac{1}{2}\left[ (\partial_\tau y(z,\tau))^2+(\partial_z y(z,\tau))^2\right]-\theta(z)(\cos y(z,\tau)-1)-\frac{4 \Gamma (\mathcal{N}_g)}{\Delta}\delta(z) (\cos y(z,\tau)-1)\right\}
\end{align}
\end{widetext}
with $y(z)=\sqrt 2 {\phi}_\sigma(z)$. It is useful to first find a classical solution for this model. The corresponding equation of motion in the region $0\ll z \ll L$ is given by
\begin{align}
\partial_{zz}y(z)-\sin y(z)=0.
\end{align}
The trivial solution for $y(z)=0$ corresponds to an even-charge state (see Eq.\eqref{eq:phisigmabc1}) whereas the inhomogeneous solution ($\pi$-soliton) describes an odd-charge state, see Fig.~\ref{fig:csoliton}.  According to  Eq.~\eqref{eq:phisigmabc2}, $y(0) \rightarrow \pm \pi$ in the middle of an odd-charge plateau. Close to the even-odd transition point, the inhomogeneous solution can be approximately written as $y(z)\approx y_0 \exp(-z)$ with $y_0 \ll 1$. Using the above inhomogeneous solution, one can write the energy functional $\mathcal F[y_0]$ in terms of the field at the boundary $y_0$:
\begin{align}\label{eq:functional}
\mathcal F[y_0]=\frac{1}{2\pi} \left(-\frac{\eta}{2}y_0^2+\frac{y_0^4}{32}\right),
\end{align}
where $\eta=4\left[\Gamma(\mathcal{N}^*_g)-\Gamma(\mathcal{N}_g)\right]/\Delta \sim \sqrt{E_Cr^2/\Delta} (\mathcal{N}_g-\mathcal{N}^*_g)$ is the detuning from the charge degeneracy point $\mathcal{N}^*_g$. Thus, for $\eta <0$ (even charge) we find that $y_c=0$ minimizes the energy functional whereas $y_c=\pm \sqrt{8 \eta}$ for $\eta >0$ (odd-charge). The two degenerate configurations correspond to the opposite spin densities (i.e. a quasiparticle with spin-up or spin-down). Thus, at the classical level, the even-odd charge transition is reminiscent of the Landau mean-field theory of second-order phase transitions. Our next step is to consider quantum fluctuations which, as we show below, destroy the quantum phase transition. Indeed, quantum fluctuations lead to tunneling between the minima of the potential \eqref{eq:functional}.

In order to investigate the nature of the quantum phase transition, we study the following effective model:
\begin{align}\label{eq:app:action_q}
S_{\rm eff}=&\frac{1}{8 \pi^2} \int d\tilde{\tau}\int d\tilde{\tau'} \frac{[y_0(\tilde{\tau})-y_0(\tilde{\tau'})]^2}{(\tilde{\tau}-\tilde{\tau'})^2} \\
&+\frac{1}{4\pi}\int d\tilde{\tau}\left(\frac{1}{2} [\dot{y_0}(\tilde{\tau})]^2 -\eta y_0^2(\tilde{\tau})+\frac{y_0^4(\tilde{\tau})}{16}\right).\nonumber
\end{align}
Here the first term originates from the bulk modes in the normal part of the nanowire (i.e. for $x<0$). We will refer to it as a dissipative term~\cite{LeggettRMP}. Without dissipation, tunneling between the two minima is described by an instanton configuration corresponding to a kink (or antikink): $y_0(\tilde{\tau}=-\infty)=-|y_c|$ and $y_0(\tilde{\tau}=\infty)=|y_c|$. The amplitude $A$ for a such process, which corresponds to a spin-flip electron backscattering and conserves the charge on the island, can be evaluated using the semi-classical approximation:
\begin{align}\label{eq:A}
\tilde{A} \equiv \frac{A}{\Delta}  \sim \exp(-S_{\rm WKB}),
\end{align}
where  $S_{\rm WKB}$  is the WKB action corresponding to a single kink (or an anti-kink). In the absence of the dissipation, kinks and anti-kinks are non-interacting. The dissipation introduces logarithmic interactions between kinks and anti-kinks which renormalizes $\tilde{A}$. In the so-called kink approximation $y_0(\tilde{\tau})=2|y_c| \sum_i \epsilon_i \delta(\tilde{\tau}-\tilde{\tau}_i)$ with $\epsilon_i=\pm 1$ and $\tau_i$ being the position of the kink. It is now straightforward to derive RG equations for $\tilde{A}$ and the strength of effective interaction between kinks and anti-kinks $\alpha$~\cite{Anderson'70, Chakravarty'82, Bray'82}:
 \begin{align}
 \frac{d\tilde {A}}{dl}&=\left(1-\frac{\alpha}{2}\right) \tilde {A}, \\
 \frac{d\alpha}{dl}&=-\alpha \tilde {A}^2.
 \end{align}
Here the initial values are $\alpha(0)=2y_c^2/\pi^2 $ and $\tilde {A}(0)$ given by Eq.\eqref{eq:A}. Since the maximum value of $y_c$ is restricted by $y_0 \leq \pi$, the initial value $\alpha(0) \leq 2$. In this regime, given that $\alpha(l)$ is decreasing under RG, the amplitude $\tilde{A}$ is increasing and there is no phase transition into a localized phase. In other words, the system is on the delocalized side of the phase transition~\cite{Anderson'70, Chakravarty'82, Bray'82}. As a result, the dependence of the ground-state energy near even-odd degeneracy point is an analytic function of $\mathcal{N}_g$. This conclusion can be also understood in terms of the single-channel antiferromagnetic Kondo problem using the mapping $(1-\alpha/2)\rightarrow J_z >0$ and $\tilde{A} \rightarrow J_{\perp}$. The latter has Fermi liquid description which corroborates our conclusion regarding the analytic dependence of observable quantities on $\mathcal{N}_g$.

\section{Coulomb blockade in the presence of Majorana states}\label{sec:spinless}

In this section, we study charging effect in a proximitized nanowire which is driven  by a magnetic field into a topologically nontrivial state and is in contact with a normal lead. Coulomb blockade in such a system has been recently studied experimentally~\cite{albrecht2015}. Application of a sufficiently strong magnetic field $B> B_c$ along the nanowire results in a topological transition~\cite{Lutchyn10, Oreg10} with Majorana zero-energy states emerging at the ends of the proximitized nanowire.

In the absence of charging energy, the presence of the zero-energy states results in the degeneracy between the states of the proximitized nanowire with even and odd number of electrons. The finite charging energy results in an $e$-periodic Coulomb blockade~\cite{Fu2010}, which is qualitatively different from the case of a conventional (e.g. s-wave) superconducting state in the wire as well as from the Coulomb blockade in a normal wire. Coupling of the proximitized wire to a normal lead broadens the Majorana resonance and leads to a continuous variation of the charge with the gate voltage, cf. Eq.~\eqref{eq:charge}. Magnetic field breaks the spin symmetry and at large fields will drive the system into a spinless regime. In the presence of charging energy, the problem reduces to that of a non-degenerate localized state broadened by coupling to a Fermi sea.


\subsection{Coulomb blockade in the weak tunneling limit}

In the case of small conductance, $g\ll 1$, the zero-energy state is broadened into a Breit-Wigner resonance of a width $g\Delta/(8\pi)$~\cite{vanHeck'16}. The Friedel sum rule applied to the resonance yields a broadened step in charge,
\begin{equation}
\frac{Q({\cal N}_g)}{e}=\frac{1}{\pi}\arctan\left\{\frac{E_C[{\cal N}_g-(n+1/2)]}{g\Delta_P/(16\pi)}\right\}+\frac{1}{2},
\label{eq:resonance}
\end{equation}
Here $g\Delta_P/16\pi \ll E_C$ with $\Delta_P$ being the p-wave gap and $n \in \mathbf{Z}$ corresponds to $n$-th step. The applicability condition of~\eqref{eq:resonance} breaks down
if the charging energy is small enough, the reason being that it was derived under the assumption that only two charge states are relevant. To go beyond this, it is convenient to cast the problem in terms of bosonic variables, and use the framework of the RG technique.

The effective Hamiltonian for the system, written in bosonic variables, is given by
\begin{align}
H&=H_{\rm W}+H_C+H_P+H_B, \label{eq:p-waveH} \\
H_{\rm W}&=\frac{v}{2\pi}\int_{-\infty}^{L}dx \left[(\partial_x \theta)^2+(\partial_x \phi)^2\right],\\
H_C&=E_c(\hat{N}-\mathcal{N}_g)^2=E_C\left(\frac{\phi(0)}{\pi}-\mathcal{N}_g\right)^2,\\
H_P&=-\frac{\Delta_P}{2\pi a}\int_{0}^{L} dx \cos(2\theta).
\end{align}
The notation here is as in Sec.~\ref{sec:strong}, except there is only a single species of bosons, and $\Delta_P$ is the induced p-wave superconducting gap, which depends on the strength of coupling between the nanowire and the s-wave superconductor, the spin-orbit coupling $\alpha$, and on the Zeeman splitting $V_Z=\mu_B g_B B$~\cite{Lutchyn10, Oreg10} (here $\mu_B$ and $g_B$ are respectively the Bohr magneton and the g-factor in the nanowire). As in Sec. \ref{sec:strong}, we assume here that the gap in the bulk superconductor $\Delta_{\rm Al}$ is large $\Delta_{\rm Al} \gg \Delta_P, E_C$, and the effective Hamiltonian~\eqref{eq:p-waveH} is valid for energies below $\Delta_{\rm Al}$, in which case the bulk superconducting degrees of freedom are frozen out. As in the previous Section, we assume that the normal-state level spacing in the nanowire $\sim v/L\to 0$.

In the weak tunneling limit, a high barrier at $x=0$ pins the field $\phi(0)$ at $\phi(0)=\pi N$ ($N \in \mathbb{Z}$). In this case, we may write the backscattering term $H_B$ in \eqref{eq:p-waveH} as
\begin{align}
H_B= D t(D)\cos\left(\theta^-(0)-\theta^+(0)\right),
\label{eq:H_B-tunn}
\end{align}
where the fields $\theta^{\pm}(\tau, x)$ are defined for $x \in [-\infty, 0]$ and  $x \in [0, L]$, respectively.  The tunneling amplitude $t_0=t(D_0)$ at the initial value of the bandwidth, $D=D_0$, must be tuned to produce the observable value of the normal-state conductance $g$ in absence of charging energy.

We are interested in extending \eqref{eq:resonance} and deriving estimates valid at $E_C, g\Delta_P\ll\Delta_P$. Evaluation of the scaling dimension of the operator $H_B$ of \eqref{eq:H_B-tunn} shows that the tunneling constant $t$ is marginal for $D>\Delta_P$, and does not flow until the bandwidth reaches $D\sim \Delta_P$. At smaller bandwidth, the proximity-induced term becomes large and gaps out bulk modes, i.e. $\theta^+(\tau, x)=\pi m$ with $m \in Z$. As a result, the scaling dimension of the tunneling Hamiltonian
\begin{align}\label{eq:dual}
H_B=D t(D)\cos \theta^-(0)
\end{align}
changes, and the RG flow at $D<\Delta_P$ becomes
\begin{align}
\frac{d t}{dl}=\frac{t}{2}.
\end{align}
One can recognize the similarity of the effective boundary term~\eqref{eq:dual} with the Majorana coupling discussed in Refs.~[\onlinecite{Fidkowski2012, Lutchyn2013}]. Since $t$ is relevant and, in the absence of charging energy $E_C=0$), would flow to strong coupling according to $t(D)=t_0 \sqrt{\Delta_P/D}$, corresponding to the so-called Andreev fixpoint. The scale $D_c$ at which the boundary term reaches strong coupling ({\it i.e.}, $t(D_c)\sim 1$) is
\begin{align}
D_c=|t_0|^2 \Delta_P \sim g \Delta_P\,.
\label{eq:Dc}
\end{align}
At this scale, the field $\theta^-(\tau, 0)$ becomes pinned and, thus, the boundary conditions for the lead electrons at $x=0$ crossover from {\it perfect} normal reflection in the ultraviolet (i.e. $\psi_R(0)=\psi_L(0)$) to {\it perfect} Andreev reflection in the infrared (i.e. $\psi_R(0)=\psi^\dag_L(0)$). One can interpret the scale $D_c$ as the broadening scale $\Gamma=g\Delta_P/(8\pi)$ in the non-interacting Majorana problem \cite{vanHeck'16}.

With finite charging energy, $E_C \neq 0$, the physics at the boundary depends on the comparison of $D_c$ and $E_C$.

If $\Gamma \sim D_c \ll E_C$, the flow of the coupling $t(D)$ is cut off by the charging energy. Therefore, away from the charge degeneracy points, $|{\cal N}_g-(n+1/2)|\gtrsim g\Delta_P/E_C$, the amplitude $t(D)$ does not reach strong coupling. Thus, the dependence of the ground-state energy on the gate voltage is not renormalized (apart from smearing of the singularities at the charge-degeneracy points, see the Eq.~\eqref{eq:resonance}) and is given by the bare charging energy
\begin{align}\label{eq:charging1}
\delta E_{\rm GS}(\mathcal{N}_g)=\min_{\mathcal{N} \in \mathbb{Z}} E_C (\mathcal{N}-\mathcal{N}_g)^2.
\end{align}

In the other case $\Gamma\sim D_c \gg E_C$, the tunneling amplitude does reach the strong-coupling limit in the entire range of gate voltages.
To proceed the renormalization for energies below $D_c$ in this situation, we switch from the tunneling Hamiltonian~\eqref{eq:H_B-tunn} to the dual description by the Hamiltonian for weak backscattering,
\begin{equation}
H_B =-D r\cos 2\phi(0).
\label{eq:backscattering_pwave}
\end{equation}
The two description should match each other at roughly $D\sim D_c$ with
\begin{equation}\label{eq:match}
 r(D_c)\sim t(D_c).
\end{equation}
{}

Now, since $\theta^-(\tau, 0)$ is pinned at low energy scales, we rewrite the boundary action for $D\lesssim D_c$ in terms of the dual fluctuating variable $\phi^-(\tau, 0)$. After integration out the $x<0$ degrees of freedom, the effective boundary theory becomes
\begin{align}\label{eq:boundary1}
S=&\frac{1}{2\pi}\!\int_0^{D_c} \!\! \frac{d\omega}{2\pi}|\omega| |\phi^-(\omega, 0)|^2\!+\!E_C\int_{{D_c}^{-1}}^{T^{-1}} \!\!\!\!\!d\tau \!\!\left(\!\!\frac{\phi^-(\tau, 0)}{\pi}\!-\!\mathcal{N}_g\!\right)^2\nonumber\\
&- D r (D)\int_{{D_c}^{-1}}^{T^{-1}} d\tau \cos 2 \phi^-(\tau, 0).
\end{align}
The boundary term proportional to $r$ is dual to the one in Eq.~\eqref{eq:dual}, and flows under RG according to
\begin{align}
\frac{dr}{dl}=-r.
\end{align}
Thus the backscattering at the junction becomes irrelevant in the RG sense for $D>E_C$, in contrast to the normal island case where it is marginal~\cite{Flensberg'93,Matveev'95}. As shown below, this change in the scaling dimension of $r$ leads to additional suppression of the charge oscillations with $\mathcal{N}_g$.

\begin{figure}
\includegraphics[width=3.5in]{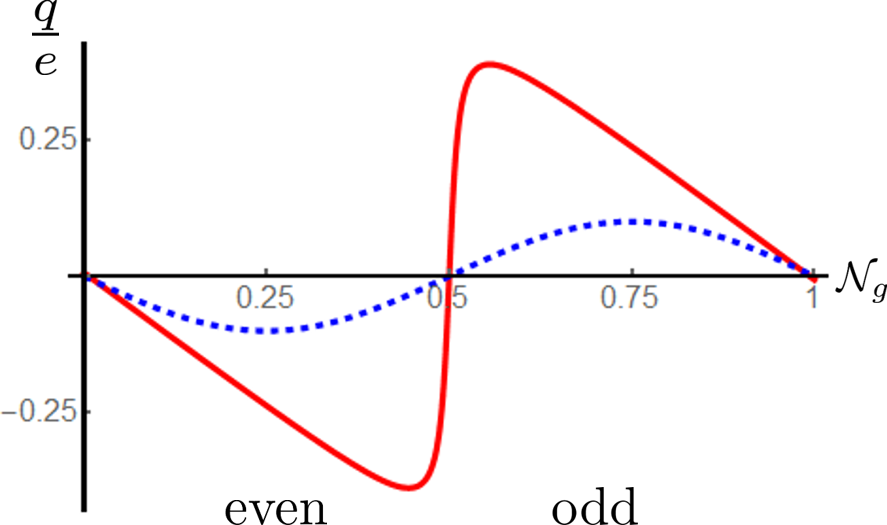}
\caption{Schematic plot of a reduced charge of a spinless proximitized nanowire as a function of the dimensionless gate voltage $\mathcal{N}_g$. The solid red and dashed blue lines correspond to weak and strong tunneling regimes, respectively.}
\label{fig:Majorana}
\end{figure}

To continue the RG procedure, we shift $\phi^-(\tau, 0) \rightarrow \phi^-(\tau, 0)+\pi \mathcal{N}_g$ and run the RG until $D\sim E_C$ to find the following boundary action:
\begin{widetext}
\begin{align}\label{eq:boundary2}
S=\frac{1}{2\pi}\int_0^{E_C} \frac{d\omega}{2\pi}|\omega| |\phi^-(\omega, 0)|^2- E_C r (E_C)\int_{{E_C}^{-1}}^{T^{-1}} d\tau \cos \left(2 \phi^-(\tau, 0)+2\pi \mathcal{N}_g\right)+\frac{E_C}{\pi^2} \int_{{E_C}^{-1}}^{T^{-1}} d\tau [\phi^-(\tau, 0)]^2,
\end{align}
\end{widetext}
where the coupling $r (E_C)\sim({E_C}/{D_c})r(D_c)\sim E_C/(g\Delta_P)$; we used Eqs.~(\ref{eq:Dc}) and (\ref{eq:match}) here. The charging energy term in Eq.~\eqref{eq:boundary2} pins the field at the boundary, $\phi^-(0)=0$. Upon substituting that value in the second term of action~\eqref{eq:boundary2}, and expressing $r (E_C)$ in terms of the bare parameters, we read off the gate voltage dependence of the ground state energy:
\begin{equation}
\delta E_{\rm GS}(\mathcal{N}_g)=-E_C^*\cos \left(2\pi \mathcal{N}_g\right)\,,\,\,\, 
E_C^* \sim \frac{E^2_C}{g\Delta_P }
\label{eq:charging3}.
\end{equation}
The crossover between the limits of~\eqref{eq:charging1} and \eqref{eq:charging3} occurs at $E_C \sim g\Delta_P$. The dependence of a reduced charge of a proximitized nanowire on gate voltage in weak and strong tunneling regimes is shown in Fig.~\eqref{fig:Majorana}.

\subsection{Coulomb blockade in the strong tunneling limit}

We now study the high-conductance limit and show that effective charging energy will be even further suppressed by quantum charge fluctuations. We concentrate first on the limit $\Delta_P \gg E_C$, in which superconductivity significantly modifies the charging effect. Assuming that the backscattering is weak, we may use the Hamiltonian~\eqref{eq:backscattering_pwave} to describe it. The backscattering term~\eqref{eq:backscattering_pwave} is marginal for $D> \Delta_P$ and does not flow until $D \sim \Delta_P$. The boundary action at a smaller bandwidth, $D \ll \Delta_P$, reads
\begin{align}\label{eq:boundary2b}
S&=\frac{1}{2\pi}\int_0^{D} \frac{d\omega}{2\pi}|\omega| |\phi^-(\omega, 0)|^2+\frac{E_C}{\pi^2} \int_{{D}^{-1}}^{T^{-1}} d\tau \phi^-(\tau, 0)^2 \nonumber\\
&- \int_{{D}^{-1}}^{T^{-1}} d\tau D r (D)\cos \left[2 \phi^-(\tau, 0)+2\pi \mathcal{N}_g \right]\,;
\end{align}
the RG flow for the coupling $r$ in the domain $\Delta_P \gg D \gg E_C$ is given by
\begin{align}
\frac{dr}{dl}=-r.
\end{align}
Following the analysis in the previous section, one finds that gate voltage dependent part of the ground-state energy is given by
\begin{align}
\delta E_{\rm GS}(\mathcal{N}_g)\sim-r\frac{E^2_C}{\Delta_P} \cos \left(2\pi \mathcal{N}_g\right)\, , r\sim \sqrt{1-g}.
\label{eq:small_r}
\end{align}
Note that the only difference with respect to weak tunneling limit (cf. Eq.\eqref{eq:charging3}) is the appearance of the bare reflection amplitude $r \ll 1$ rather than $r(D_c)\sim 1$. Therefore, the effective charging energy vanishes for $r \rightarrow 0$.

Finally, we note that when $E_C \gg \Delta_P$, the effect of the superconductivity is negligible; the effective charging energy follows from Refs.~\cite{Flensberg'93,Matveev'95}
\begin{align}
\delta E_{\rm GS}(\mathcal{N}_g)\sim- E_C r \cos \left(2\pi \mathcal{N}_g\right).
\label{eq:small_r1}
\end{align}
The two equations \eqref{eq:small_r} and \eqref{eq:small_r1} match at $E_C\sim\Delta_P$.

\section{Conclusions}\label{sec:conclusions}

In this paper, we have studied charging effects of a proximitized nanowire in contact with a normal lead. We considered two different regimes which were recently investigated experimentally~\cite{albrecht2015}: (a) the spinless case, emerging when the nanowire is driven into topological superconducting state (i.e. $B > B_c$ with $B_c$ being the critical field corresponding to the topological phase transition~\cite{Sau10, Lutchyn10, Oreg10}) and (b)  the spinful case (zero magnetic field $B=0$). In both these cases, we calculate the charge of the proximitized nanowire $Q$ as a function of the dimensionless gate voltage ${\cal N}_g$.

The main difference of the charge staircase for a Majorana Coulomb island as compared to a normal-state island is in the step width. At the same small conductance $g$ of the junction, the step width is much larger in the former system and it scales as $\propto g$ (see Eq. \eqref{eq:resonance}) as opposed to $\propto\exp\left(-\frac{\pi^2}{2\sqrt{g}}\right)$ in the normal-state case~\cite{Matveev'95, Furusaki'95}. Further differences come with an increase of the junction conductance. In either system, the steps are progressively washed out with increasing $g$ and vanish as $g \rightarrow 1$. However, for a Majorana Coulomb island the steps degrade with the increase of $g$ faster than for a normal-state island: in the case of large p-wave gap $\Delta_P\gg E_C$, the crossover from charge steps to a weak harmonic modulation of charge occurs at $g\sim E_C/\Delta_P\ll 1$~\eqref{eq:charging3}, while such crossover requires $g\sim 1$ in the normal-state case~\cite{Matveev'95}.

In the spinful case, the charge staircase for a single-channel junction was investigated in great detail for the case of a normal-state Coulomb island~\cite{Aleiner'02}. It turns out that the function $Q({\mathcal N}_g)$ is non-analytic at half-integer values of ${\mathcal N}_g$ for any value of $g$ in the normal case. The non-analyticity at the charge-degeneracy points stems from the equivalence of the Coulomb blockade problem to a version of a symmetric two-channel Kondo problem: two electron spin states in the former problem map on the two channels in the latter one. However, much less was known about the shape of the steps in the case of superconducting island, even for the $s$-wave superconductor. The case $\Delta>E_C$, $g\ll 1$ for an $s$-wave superconducting island was considered in Ref. \cite{Garate'11}. It was shown there that the $2e$-charge degeneracy points occurring at odd integer values of ${\mathcal N}_g$ are described by a single-channel ``charge-Kondo'' model, thus $Q({\mathcal N}_g)$ is analytic across the charge degeneracy point. In this work, we demonstrated that the analyticity of $Q({\mathcal N}_g)$ is preserved for any values of conductance $g$, regardless of the ratio $\Delta/E_C$. In particular, we showed that the sharp charge-$e$ steps which occur at $g\rightarrow 0$  and $\Delta < E_C$ are broadened and are described by an analytic function of ${\mathcal N}_g$ .

Considering the case of a large conductance, we found how the charge staircase is smeared out with $g \rightarrow 1$, see Eqs.~\eqref{eq:charging_spinful1},\eqref{eq:charging_spinful2}, \eqref{eq:eg_even} and \eqref{eq:eg_odd}. We also identified the analogue of the even-odd transitions earlier known to occur in the case $g\to 0$, if $\Delta<E_C$. The corresponding condition at high conductance $g$ involves a charging energy which is renormalized by quantum fluctuations to $E^*_C\sim E_C (1-g)$. We have shown that the even-odd transition is actually a crossover as a function of ${\mathcal N}_g$, see the discussion after Eq. \eqref{eq:app:action_q}, and found the crossover width, see Eqs.\eqref{eq:eg_even} and \eqref{eq:eg_odd}.

Thus, the present theory of Coulomb blockade in proximitized wires spans the limits of low and high junction conductance and applicable for $s$- and $p$-wave superconductors. It may open ways to detect Majorana states in high-precision charge measurements, and is relevant for the electrostatic manipulation of such states in the topological quantum computing proposals based on nanowire networks~\cite{Aasen2015}.

\section{Acknowledgements}

We are grateful to A. Kamenev, C. M. Marcus, and K. Matveev for discussions. This work is supported by DOE contract DEFG02-08ER46482 (LG). RL and LG acknowledge the hospitality of the Aspen Center for Physics supported by NSF grant No. PHY1066293, where part of this work was done. KF acknowledges support by the Danish National Research Foundation.

%
\end{document}